\documentclass[sort&compress]{elsarticle}

\usepackage{lineno,hyperref}
\modulolinenumbers[5]

\journal{Journal of Magnetism and Magnetic Materials}

\begin{document}

\begin{frontmatter}

\title{Guided Skyrmion Motion Along Pinning Array Interfaces}

\author[1,2]{N. P. Vizarim}
\author[1]{C. Reichhardt}
\author[3]{P. A. Venegas}
\author[1]{C. J. O. Reichhardt}
\address[1]{Theoretical Division and Center for Nonlinear Studies,
  Los Alamos National Laboratory, Los Alamos, New Mexico 87545, USA}
\address[2]{POSMAT - Programa de P{\' o}s-Gradua{\c c}{\~ a}o em Ci{\^ e}ncia e Tecnologia de Materiais, Faculdade de Ci\^{e}ncias, Universidade Estadual Paulista - UNESP, Bauru, SP, CP 473, 17033-360, Brazil}
\address[3]{Departamento de F\'{i}sica, Faculdade de Ci\^{e}ncias, Universidade Estadual Paulista - UNESP, Bauru, SP, CP 473, 17033-360, Brazil}

\begin{abstract}
We examine ac driven skyrmions interacting with the interface
between two different obstacle array structures. We consider
drive amplitudes at which skyrmions in a
bulk obstacle lattice undergo only localized motion
and show that when an obstacle lattice interface is introduced,
directed skyrmion transport can occur along the interface.
The skyrmions can be guided by a straight interface and can also
turn corners to follow the interface.
For a square obstacle lattice
embedded in a square pinning array with a larger lattice constant, we 
find that skyrmions can undergo transport
in all four primary symmetry directions under the same fixed
ac drive.
We map where localized or translating motion occurs as a function of
the ac driving parameters.
Our results suggest a new method 
for controlling skyrmion motion based on transport
along obstacle lattice interfaces.
\end{abstract}


\end{frontmatter}

\section{Introduction}
Skyrmions are particlelike magnetic textures that are
stabilized by their topological properties \cite{Muhlbauer09,Yu10,Nagaosa13}.
They have attracted growing
interest as an increasing number of materials 
have been identified in which
skyrmions are stable over an extended range of parameters, including 
at room temperature
\cite{EverschorSitte18,Tokunaga15,Woo16,Soumyanarayanan17,Legrand17,Fert17,Montoya18}.
Skyrmions also have interesting basic science properties
as a representative example of an emergent phenomenon
in which the
microscopic degrees of freedom in the form of spins
act collectively to form particle-like objects which in turn
experience competing interactions with temperature,
disorder,
other skyrmions,
and coupling to external drives.
Due to their size,
stability, and
ability to be manipulated
with external drives, skyrmions are also very promising 
for applications such as magnetic based memory and logic devices \cite{Fert13}. 
There have been a number of proposals
on how to create different types of devices for moving skyrmions
that interact with some form of nanostructured
sample or pinning array
\cite{Tomasello14,Zhang15,Fert17,Pinna18,Prychunenko18,Song20}. 

One of the properties of skyrmions that makes them
distinct from many other particlelike
systems interacting with a substrate
is that skyrmions have a strong gyroscopic component to their
motion in the form of a Magnus force \cite{Nagaosa13,Buttner15,Brown18}. 
The Magnus force generates a velocity component
that is perpendicular to the net force on the skyrmion,
and tends to produce spiraling motion in the presence of localized defects
as well as a skyrmion Hall effect in which the skyrmion
moves at an angle with respect to an external drive
\cite{Nagaosa13,Buttner15,Brown18,Reichhardt15,Jiang17,Litzius17}.
When a moving skyrmion interacts with a pinning site or barrier,
the Magnus force can strongly affect
the skyrmion motion,
resulting in a drive dependence of the skyrmion Hall angle
\cite{Reichhardt15,Jiang17,Litzius17,Woo18,Reichhardt19,Juge19,Zeissler20,Reichhardt20a},
as well as acceleration effects or skyrmion deflection
\cite{GonzalezGomez19,CastellQueralt20,Zhang15a,Xing20}. 

For skyrmions interacting with a periodic array of pinning sites
or obstacles under a dc drive,
the skyrmion Hall angle is not constant but increases with 
increasing skyrmion velocity and exhibits
a series of steps produced when
the skyrmion motion locks to symmetry directions of the substrate,
followed by a saturation at high velocities to the intrinsic or
disorder free value
\cite{Reichhardt15a,Feilhauer19,Vizarim20}.
When combined ac and dc driving is used to move
skyrmions over a periodic pinning array,
a variety of 
both directional locking and phase locking effects
appear \cite{Vizarim20a,Vizarim20b}.
The phase locking effects resemble Shapiro steps
in which the
ac drive frequency matches with the intrinsic frequency of the
motion of the skyrmion over the periodic substrate,
leading to steps in the skyrmion velocity-force curves
\cite{Shapiro63,Benz90}.
The Magnus
force can generate additional features in the Shapiro steps
that are not found in overdamped systems such as 
superconducting vortices moving
over two-dimensional (2D) periodic pinning under dc and ac driving
\cite{vanLook99,Reichhardt00b}. 

Under purely ac driving, 
skyrmions 
interacting with a 2D periodic substrate can pass through a series of localized 
and delocalized orbits as the size of the orbit increases \cite{Vizarim20c}.
In
the localized orbits,
the skyrmion encircles or moves through an integer number of pinning sites.
For ac parameters which place the skyrmion orbit
at the boundary
between two different stable commensurate orbits,
the motion is chaotic and the skyrmion diffuses through the system.
It is even possible in some
cases to obtain directed skyrmion motion under purely ac driving by
using a biharmonic or circular ac drive to induce a
ratchet effect
\cite{Vizarim20a,Vizarim20c}.  
In studies of skyrmions
interacting with periodic obstacle arrays,
the directed motion has a fixed orientation if the
ac drive amplitude and frequency is held fixed
\cite{Vizarim20c}. 
For potential device applications, it would valuable
to be able to steer the
skyrmions along a specified trajectory involving arbitrary
directions using only a single applied ac drive.

\begin{figure}
\includegraphics[width=3.5in]{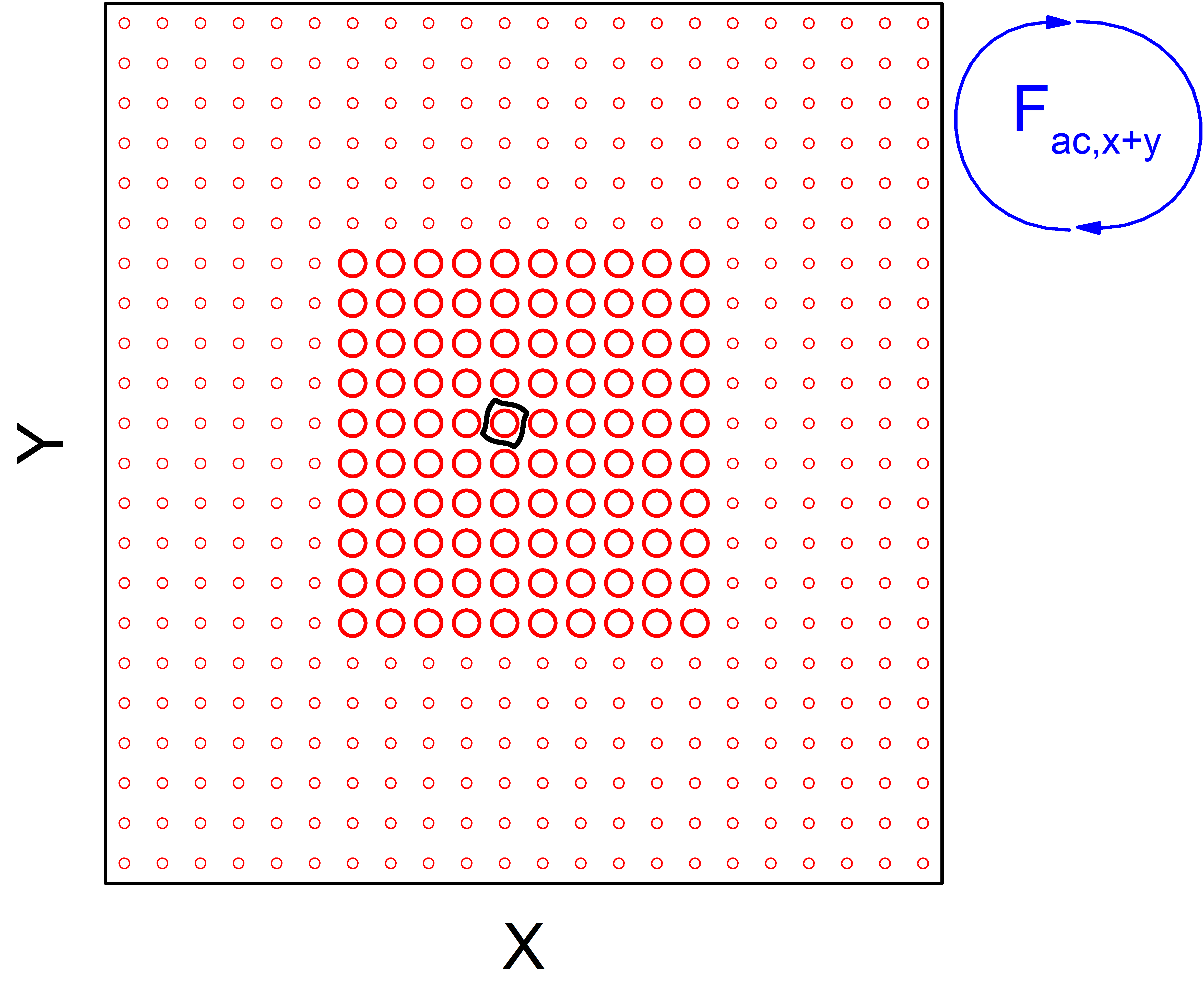}
\caption{Image of a system containing a square array of
larger obstacles (large circles) with radius $R^L_o=0.85$
set inside a square array of smaller obstacles
(small circles) with radius $R^S_o=0.45$. Both
arrays have the same lattice constant.
The black line indicates a trajectory of a skyrmion trapped in
the larger obstacle array under a circular ac drive
$F_{\rm ac, x+y}$.
}
\label{fig:1}
\end{figure}

Here we demonstrate that steerable skyrmion motion can be achieved with
a single ac drive in a system where the skyrmions
interact with an interface between two
different periodic obstacle arrays, as illustrated in Fig.~\ref{fig:1}. 
We specifically show that the skyrmion can move
along the interface and turn corners, allowing controlled motion
of the skyrmion in all four lattice symmetry directions.
The motion along the interface is quantized,
and a single interface can produce different modes of motion as
the ac drive parameters are varied.
In some cases the skyrmion remains locked along the interface and follows
the edges, while in other cases
the skyrmion can travel along a single straight interface but
decouples from the interface upon reaching a corner.
The skyrmion motion we observe is similar to the edge transport
phenomenon found
in colloidal topological insulators \cite{Loehr18},
where particle transport can occur along the interface
between two different substrate
geometries. 
Our results show
that skyrmion edge transport can be realized
and that it represents a new method for controlling
and steering skyrmion motion to create new types of devices.

\section{Simulation}

We consider a 2D system with periodic boundary conditions in
the $x$ and $y$ directions.
The sample contains
two periodic arrays of obstacles with
identical lattice constant $a$ but varied obstacle radii $R^L_o$
(for the large obstacles) and $R^S_o$ (for the small obstacles).
We introduce a single skyrmion
near the interface edge or in the bulk and drive
it with a circular ac force.
The skyrmion dynamics is modeled using the modified Thiele
equation approach as in previous work
\cite{Reichhardt15,Reichhardt20a,Reichhardt15a,Vizarim20}, 
where the equation of motion for the skyrmion is:
\begin{equation}
\alpha_d {\bf v}_{i} + \alpha_m {\hat z} \times {\bf v}_{i} =  {\bf F}^{\rm obs}_i + {\bf F}^{ac} .
\end{equation}
Here $\alpha_{d}$ is the damping term which aligns
the skyrmion velocity in the direction of the external forces,
while $\alpha_{m}$ is the Magnus force 
which creates a velocity component perpendicular to the forces on the skyrmion. 
We set ${\alpha }^2_d+{\alpha }^2_m=1$.
In the absence of pinning or obstacles and under a dc drive,
the skyrmion would move at an angle known as the skyrmion Hall angle,
$\theta_{sk}= \arctan(\alpha_{m}/\alpha_{d})$ 

The force between an obstacle and the skyrmion is given by
${\bf F}^{\rm obs}_i=-\mathrm{\nabla }U_o=-F_or_{io}e^{-{\left({r_{io}}/{a_o}\right)}^2}{\bf \hat r}_{io}\ $,
with $F_o=2U_o/R^2_o$.
The potential energy of the obstacle  
is $U_o=C_oe^{-{\left({r_{io}}/{a_o}\right)}^2}$, where $C_o$ 
is the strength of the obstacle potential, 
$r_{io}$ is the distance between skyrmion $i$ and obstacle $o$,
and $R_o=R_o^L$ or $R_o^S$ is the obstacle radius.
We fix $R_o^L=0.85$ and $R_o^S=0.45$.
For computational efficiency, we cut off
the obstacle interaction beyond $r_{io}=2.0$ since the interaction
is negligible for larger  
distances.
We set the obstacle density to
$\rho_o=0.093$
and use a total system size of
$72 \times72$.
The sample contains two different obstacle lattices which have the same 
lattice constant but different obstacle radii.
In Fig.~\ref{fig:1}
we show an example
in which the larger obstacles are embedded into a square region of the
smaller obstacle lattice.
We can also place one size of obstacle in each half of the sample to
create a pair of one-dimensional (1D) interfaces, as
in Fig.~\ref{fig:2}.
The skyrmion couples to a
biharmonic ac drive
of the form
${\bf F}^{ac} = A\sin(\omega t){\bf \hat{x}} + B\cos(\omega t){\bf \hat{y}}$,
where in this work we set $A=B$.
The ac drive frequency
$\omega$ is measured in inverse simulation time steps.
We measure the normalized average  
skyrmion velocity in the $x$ direction,
$\langle V_{x}\rangle =\langle {\bf v}_i \cdot {\bf \hat x}\rangle/\omega a$,
and in the $y$ direction,
$\langle V_{y}\rangle=\langle {\bf v}_i \cdot {\bf \hat y}\rangle/\omega a$.
Under this normalization, a value of $\langle V_x\rangle=1$
($\langle V_y\rangle=1$) indicates that the
skyrmion is moving by one lattice constant in the $x$ ($y$) direction during
each ac drive cycle.
We increase the ac drive amplitude $A$
in small steps of $\delta A=0.002$ 
and wait $10^5$ simulation time steps between increments.

\section{Transport Along a 1D Interface}

\begin{figure}
  \begin{minipage}{3.5in}
\includegraphics[width=3.5in]{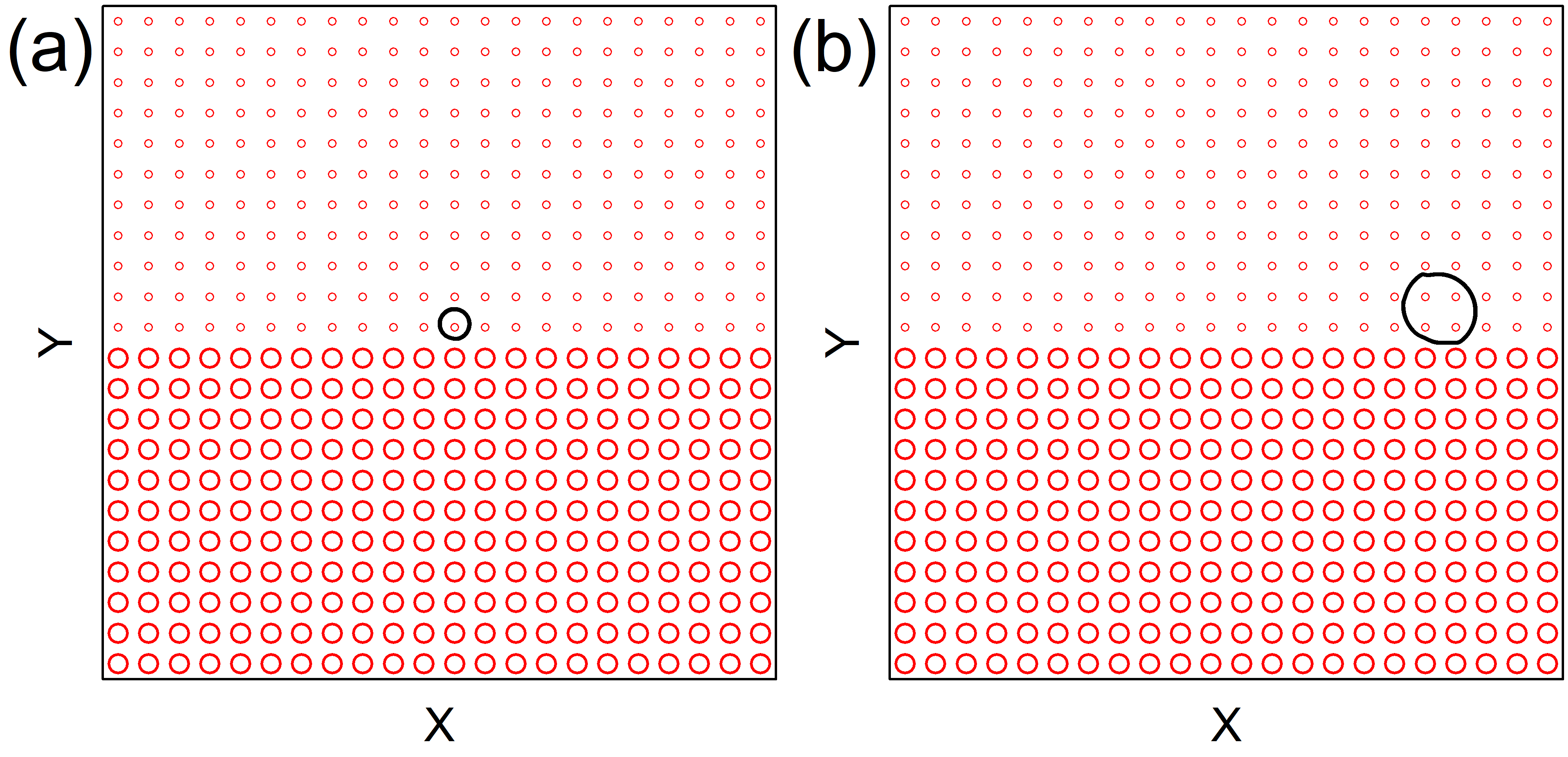}
\includegraphics[width=3.5in]{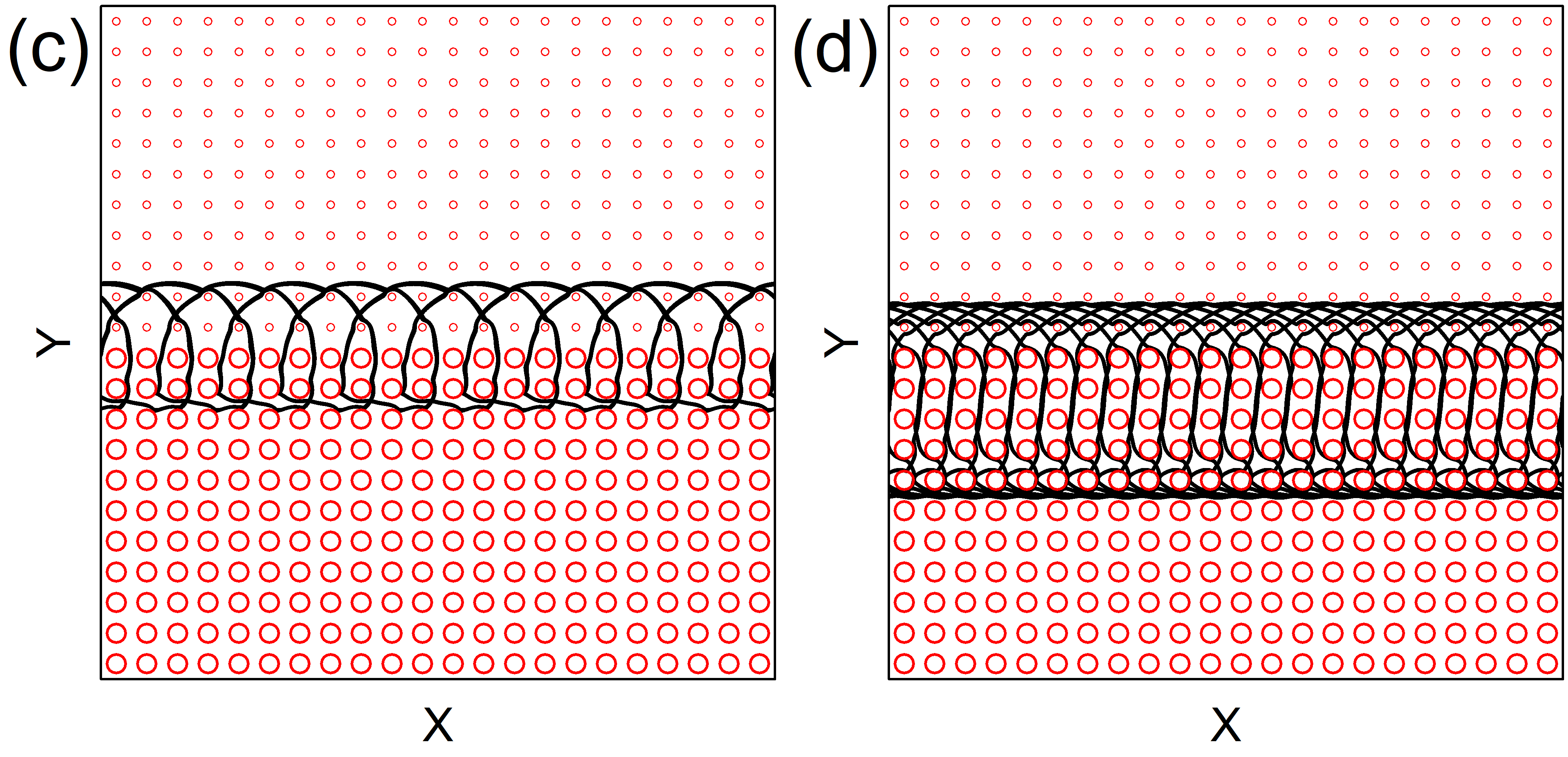}
\end{minipage}
\caption{Obstacle locations (circles) and skyrmion trajectory (lines)
in a system with a 1D interface between the large and small obstacles
at   
$\alpha_{m}/\alpha_{d} = 0.45$ under a circular ac drive of
frequency $\omega = 1\times 10^{-5}$ and amplitude $A$.
(a) At $A=0.1$ the motion is localized.
(b) A larger localized orbit at $A = 0.4$. (c) $A = 0.5$ where
the skyrmion translates in the positive $x$-direction by $2a$
during every ac cycle. 
(d) At $A = 0.77$, there is another translating orbit where the skyrmion
moves a distance $a$ in the positive $x$ direction
every ac cycle.  
}
\label{fig:2}
\end{figure}

We first consider a skyrmion interacting with a 1D interface 
as shown in Fig.~\ref{fig:2}. We can place the skyrmion within the middle of
one of the obstacle lattices
or along the interface.
In Fig.~\ref{fig:2},
the small obstacles are in the upper portion of the sample,
the large obstacles are in the lower portion of the sample,
$\alpha_m/\alpha_d=0.45$, and $\omega=1 \times 10^{-5}$.
If the skyrmion interacts with only one size of obstacle, it undergoes
either localized motion or chaotic delocalized motion but
shows little to no directed transport. 
When the skyrmion is placed near the interface for ac drive amplitude $A=0.1$,
as in Fig.~\ref{fig:2}(a),
it moves in a periodic orbit around a single obstacle.
If the particle is placed on the other side
of the interface, it again forms a localized orbit which either circles 
one obstacle or follows a closed path between adjacent obstacles.
If the drive amplitude is increased to $A=0.4$, as in
Fig.~\ref{fig:2}(b),
the skyrmion follows a localized
orbit that encircles four obstacles.
Depending on the initial placement of the
skyrmion, there can be an initial transient of chaotic motion, but
the skyrmion quickly settles into
a localized state for this value of $A$.
It is also possible for the skyrmion to
exhibit chaotic motion on one side of the interface but to settle into
a localized orbit once it migrates to the other side of the interface.
In Fig.~\ref{fig:2}(c) we show the formation of a translating orbit at $A = 0.5$
where the skyrmion moves a distance $2a$ in the positive $x$ direction
during every ac drive cycle.
Figure~\ref{fig:2}(d) illustrates
another translating orbit at $A = 0.77$,
where the skyrmion moves a distance $a$ in the positive positive $x$-direction
per ac drive cycle.

\begin{figure}
\includegraphics[width=3.5in]{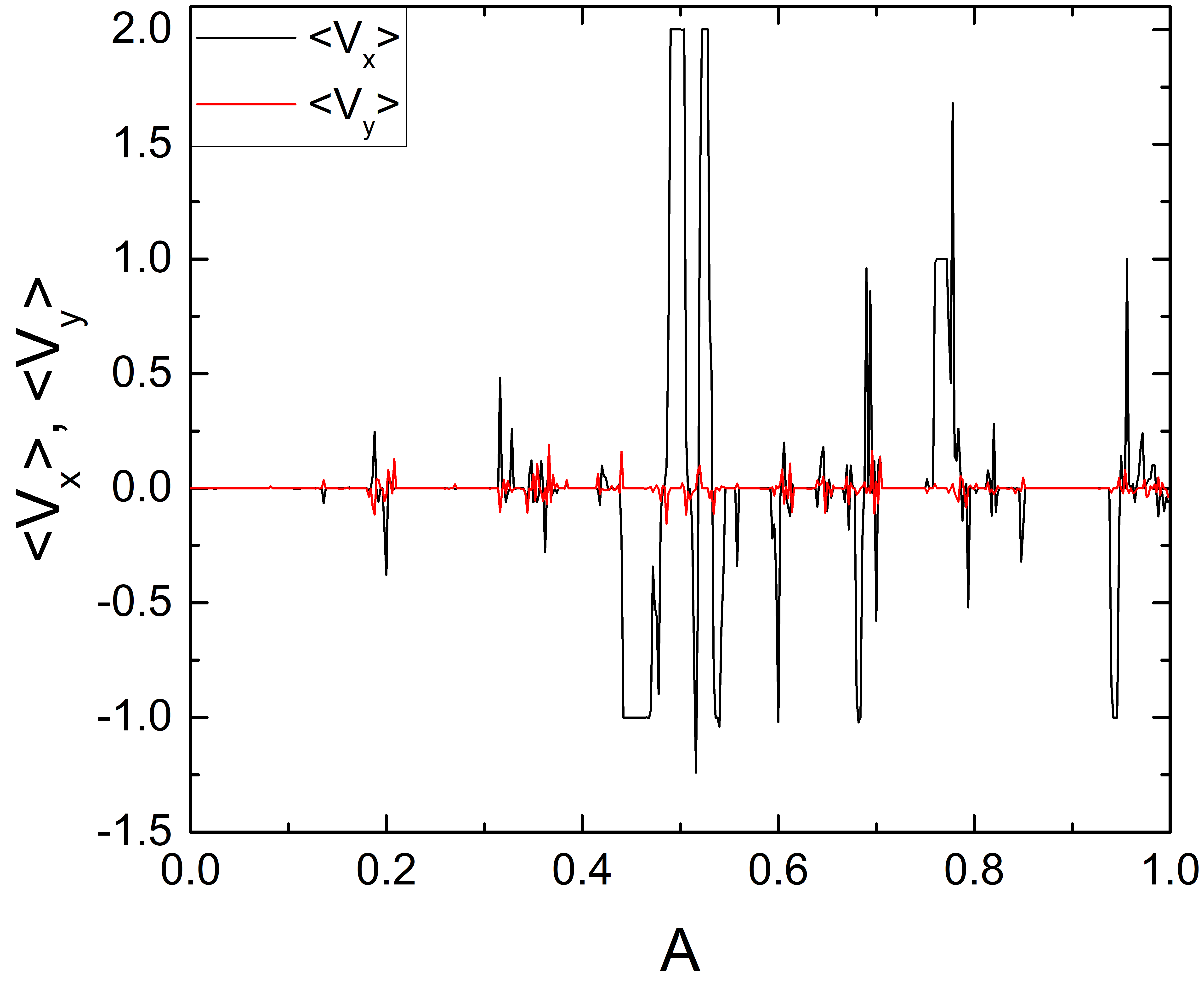}
\caption{ $\langle V_{x}\rangle$ (black) and $\langle V_{y}\rangle$ (red)
vs $A$ for the system in Fig.~\ref{fig:2}
with
$\alpha_m/\alpha_d=0.45$ and
$\omega=1 \times 10^{-5}$.
Here there are several regions where
$\langle V_{x}\rangle = \pm 1.0$ or $2.0$, indicating
translating orbits.  
}
\label{fig:3}
\end{figure}

In 
Fig.~\ref{fig:3} we plot
$\langle V_{x}\rangle$ and $\langle V_{y}\rangle$
versus ac amplitude $A$ for the system in
Fig.~\ref{fig:2}.
Here $\langle V_{y}\rangle \approx 0$ for all $A$ since
transport can occur only parallel to the interface, which is aligned
in the $x$ direction.
The skyrmion can translate in either the positive or negative $x$
direction.  When $0.425 < A < 0.75$,
the skyrmion moves along the interface
a distance $a$ in the negative $x$ direction
during each ac drive cycle,
while for $0.475 < A < 0.51$, the skyrmion translates in the positive $x$
direction by
$2a$ during each cycle.
There are also several other windows in which the skyrmion
translates by $a$ or $a/2$ in the
positive or negative $x$ direction.
If the ac drive polarity is reversed,
we obtain the same curves shown in Fig.~\ref{fig:3} but with
the $y$-axis flipped. 
The directed motion occurs along the interface due to a combination
of the broken time symmetry from the ac drive and the broken spatial
symmetry of the interface,
which produces a ratchet effect.
It was shown in previous work that when a particle
is interacting with a periodic substrate, directed motion can occur
in the absence of an interface
if the ac drive
itself breaks spatial symmetry, as can be achieved with biharmonic driving
of mixed drive amplitude or frequency
\cite{Reichhardt03}.
In the case we consider here, the ac driving is spatially symmetric,
but when one portion of the skyrmion orbit is on one side of the interface
and the other portion of the orbit is on the other side of the interface,
the orbit becomes asymmetric and the ratchet effect can occur.

\begin{figure}
\includegraphics[width=3.5in]{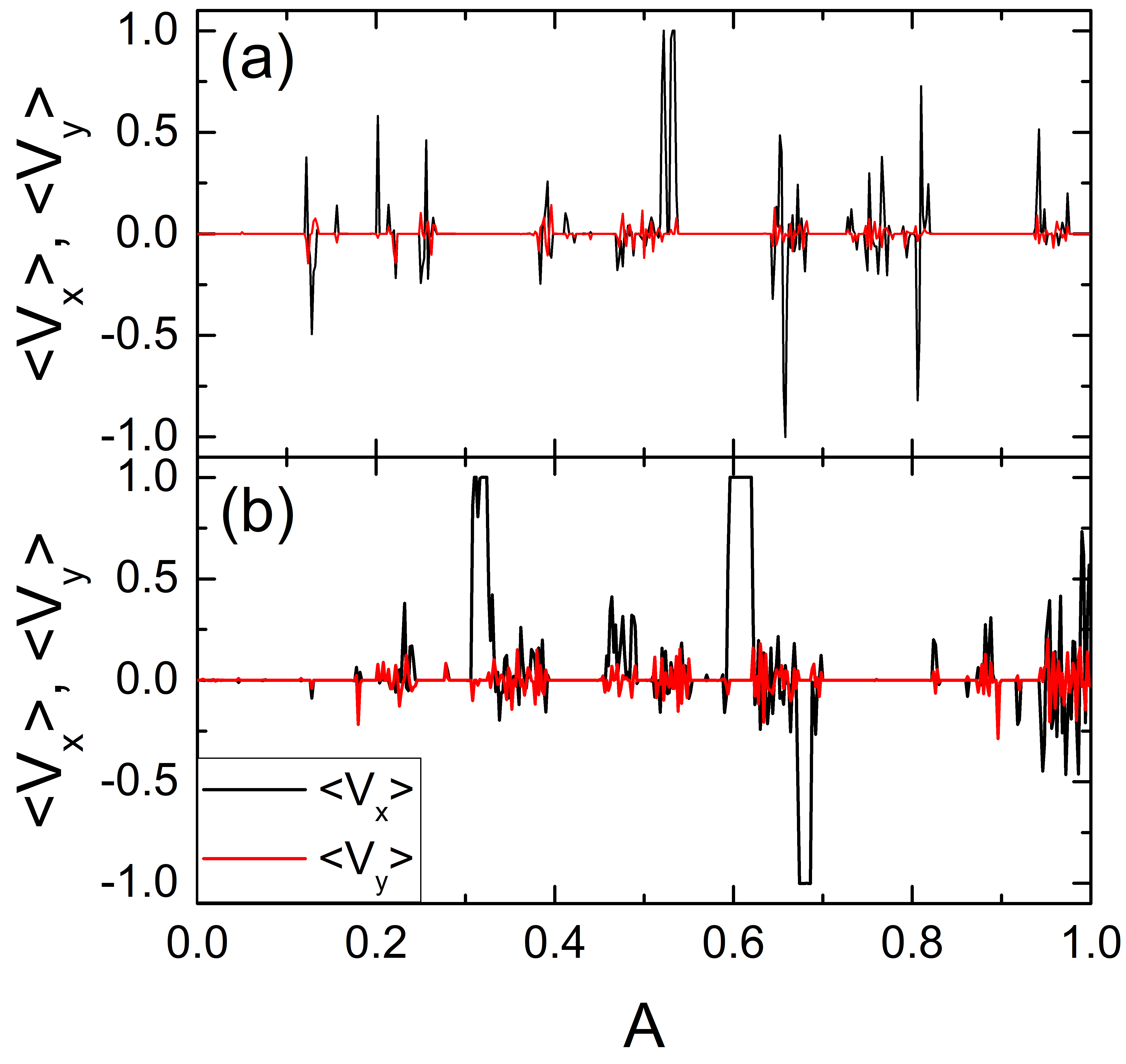}
\caption{ 
$\langle V_{x}\rangle$ (black) and $\langle V_{y}\rangle$ vs $A$
for the system in Fig.~\ref{fig:2}
with
$\omega=1\times 10^{-5}$.
(a) $\alpha_{m}/\alpha_{d} = 1.732$.
(b) $\alpha_{m}/\alpha_{d} = 9.962$, where extended
windows of directed motion appear.
  }
\label{fig:4}
\end{figure}

\begin{figure}
  \begin{minipage}{3.5in}
    \includegraphics[width=3.5in]{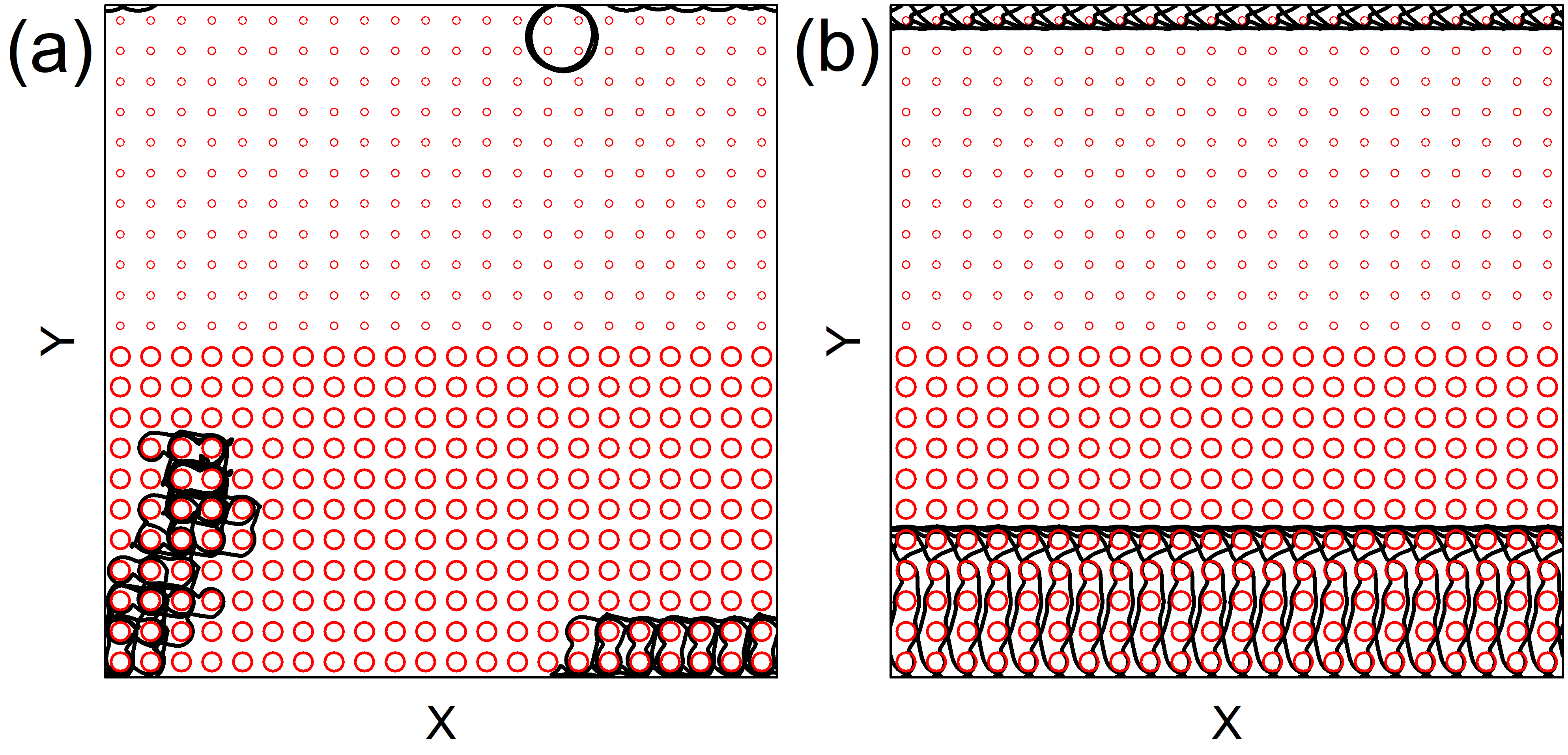}
    \includegraphics[width=3.5in]{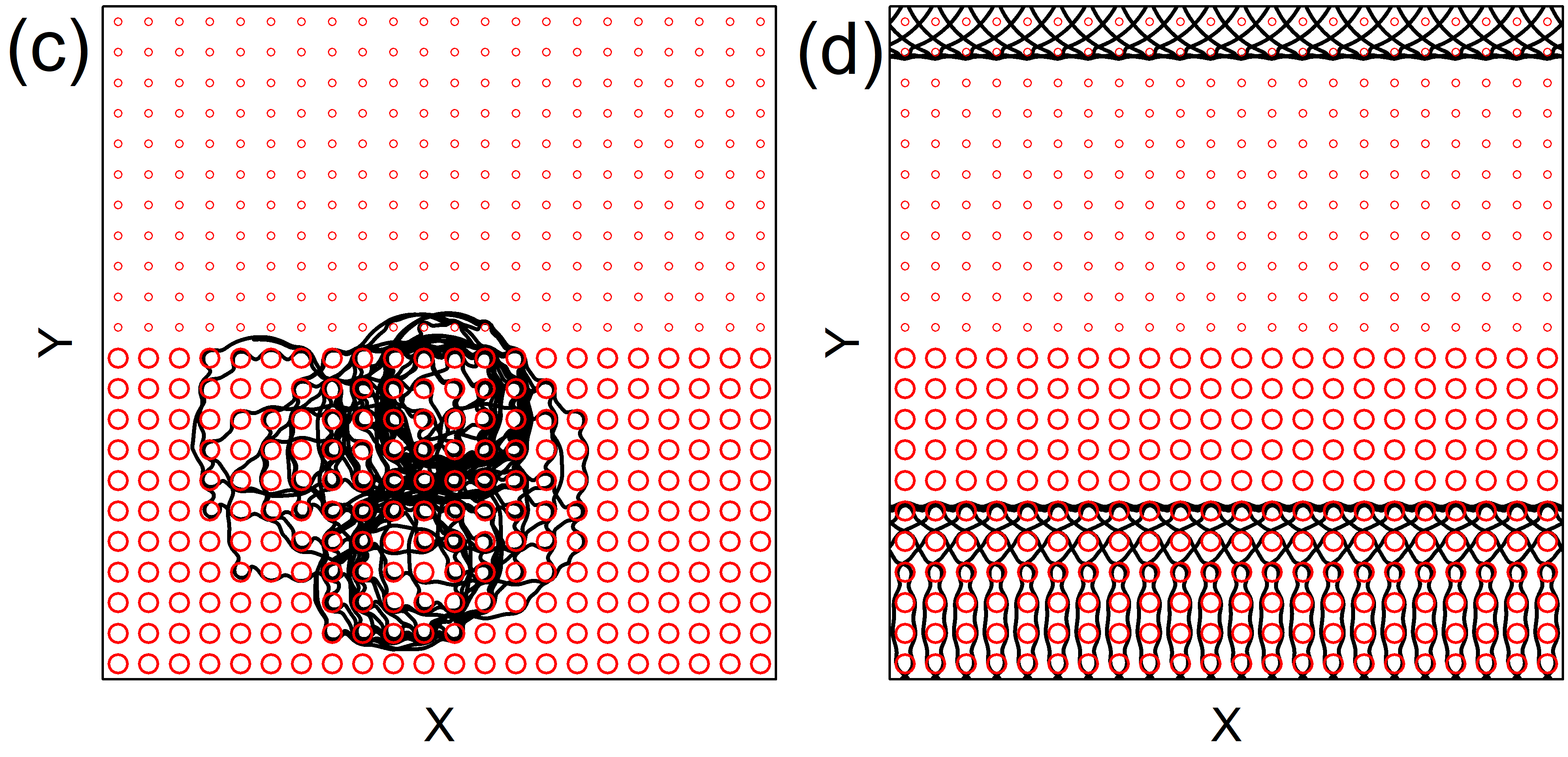}
  \end{minipage}
\caption{The obstacle locations (circles) and skyrmion trajectories (lines)
for the system in Fig.~\ref{fig:4} with
$\omega=1\times 10^{-5}$.
(a) At $\alpha_m/\alpha_d=1.732$ and $A = 0.22$
there is a short time transient of directed motion before
the skyrmion motion delocalizes.
(b) A translating orbit at $\alpha_m/\alpha_d=1.732$ and $A = 0.534$.
(c) A delocalized orbit at $\alpha_m/\alpha_d=9.962$ and $A = 0.372$.
(d) A complex translating orbit at $\alpha_m/\alpha_d=9.962$ and
$A = 0.6$.
}
\label{fig:5}
\end{figure}

In Fig.~\ref{fig:4} we plot $\langle V_{x}\rangle$ and $\langle V_{y}\rangle$
versus $A$ for the same system as in Fig.~\ref{fig:2} but with
a stronger Magnus force component of $\alpha_{m}/\alpha_{d} = 1.732$
[Fig.~\ref{fig:4}(a)] 
or $\alpha_{m}/\alpha_{d} = 9.962$
[Fig.~\ref{fig:4}(b)].
  For $\alpha_{m}/\alpha_{d} = 1.732$ in Fig.~\ref{fig:4}(a), there are 
some small regions of directed motion,
while for $\alpha_{m}/\alpha_{d} = 9.962$ in Fig.~\ref{fig:4}(b), the
directed motion regions
are more extended.
In Fig.~\ref{fig:5}(a) we illustrate the skyrmion trajectories for
the system in Fig.~\ref{fig:4}(a) at $A = 0.22$, where 
the skyrmion starts at the edge of the interface
and exhibits
a short time transient translation
in the small obstacle region
before undergoing delocalized 
motion in the large obstacle region.
In this case there is some translation in the positive $x$ direction but the
overall motion is a mixture of localized, chaotic and translating orbits. 
The skyrmion eventually becomes localized in the small obstacle region. 
Figure~\ref{fig:5}(b) shows the trajectory for the same system at
$A = 0.534$, where the skyrmion forms a translating orbit moving
a distance $a$ in the positive $x$ direction during each drive cycle.
In Fig.~\ref{fig:5}(c) we plot a delocalized orbit for the system in
Fig.~\ref{fig:4}(b) with $\alpha_m/\alpha_d=9.862$ at $A = 0.382$, while 
Fig.~\ref{fig:5}(d) shows the same system at
$A = 0.6$ were the skyrmion moves in a complex orbit at the edge of
the interface that translates a distance $a$ in the positive $x$ direction
during each ac cycle.
In general, as $\alpha_{m}/\alpha_{d}$ increases, the
non-translating states take the form of
chaotic or delocalized paths rather than closed localized orbits.  

\section{Guided Transport Along Corners}

\begin{figure}
  \begin{minipage}{3.5in}
\includegraphics[width=3.5in]{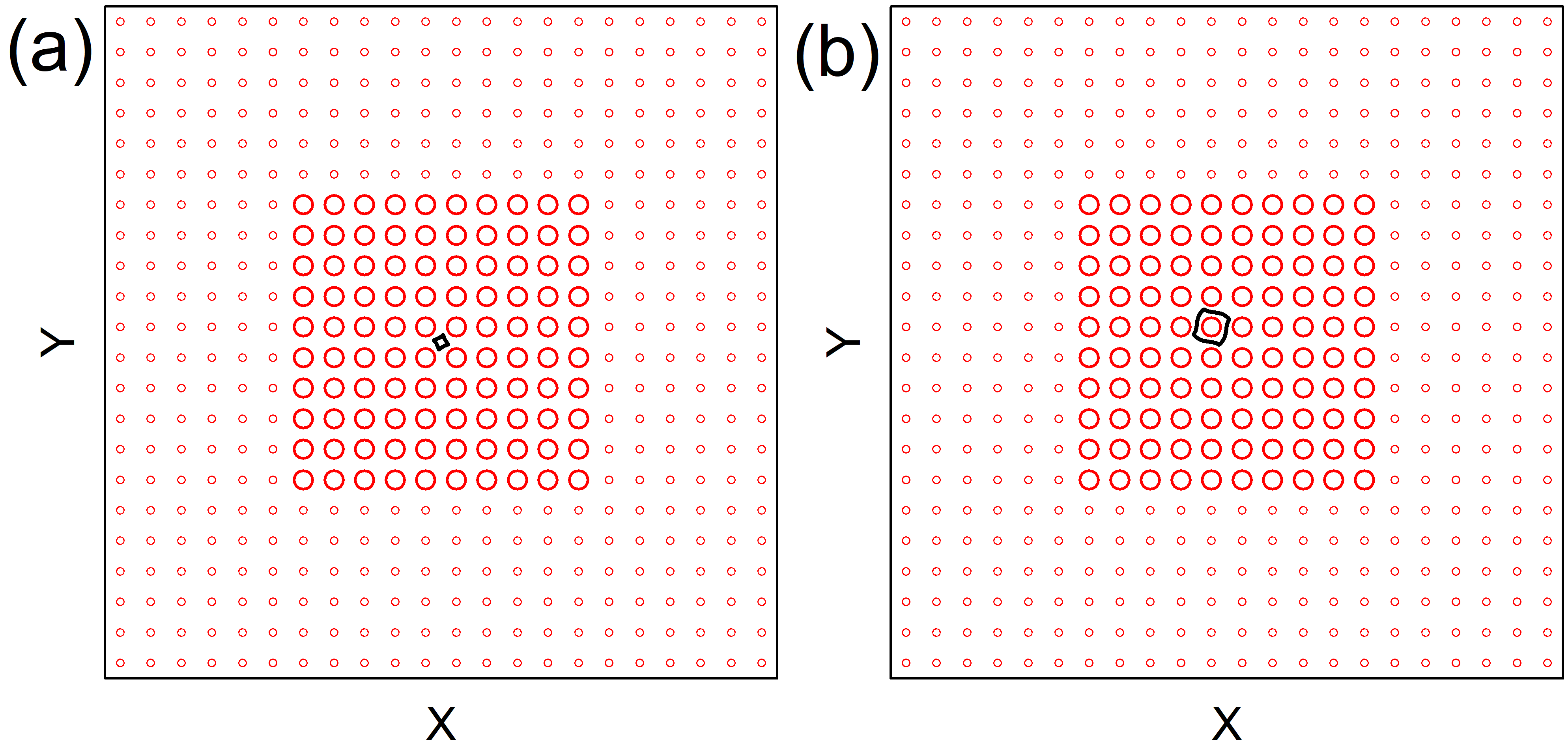}
\includegraphics[width=3.5in]{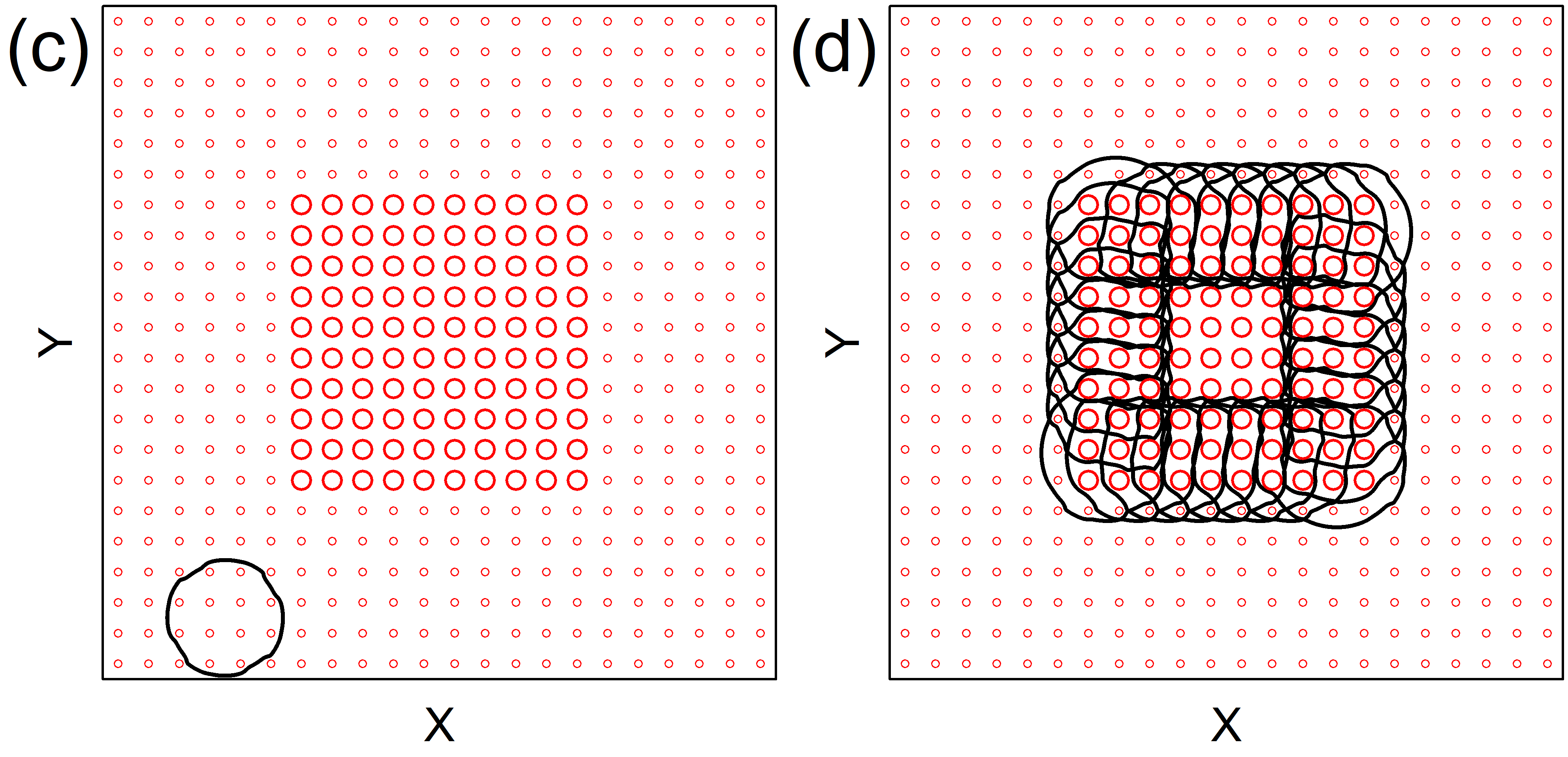}
\includegraphics[width=3.5in]{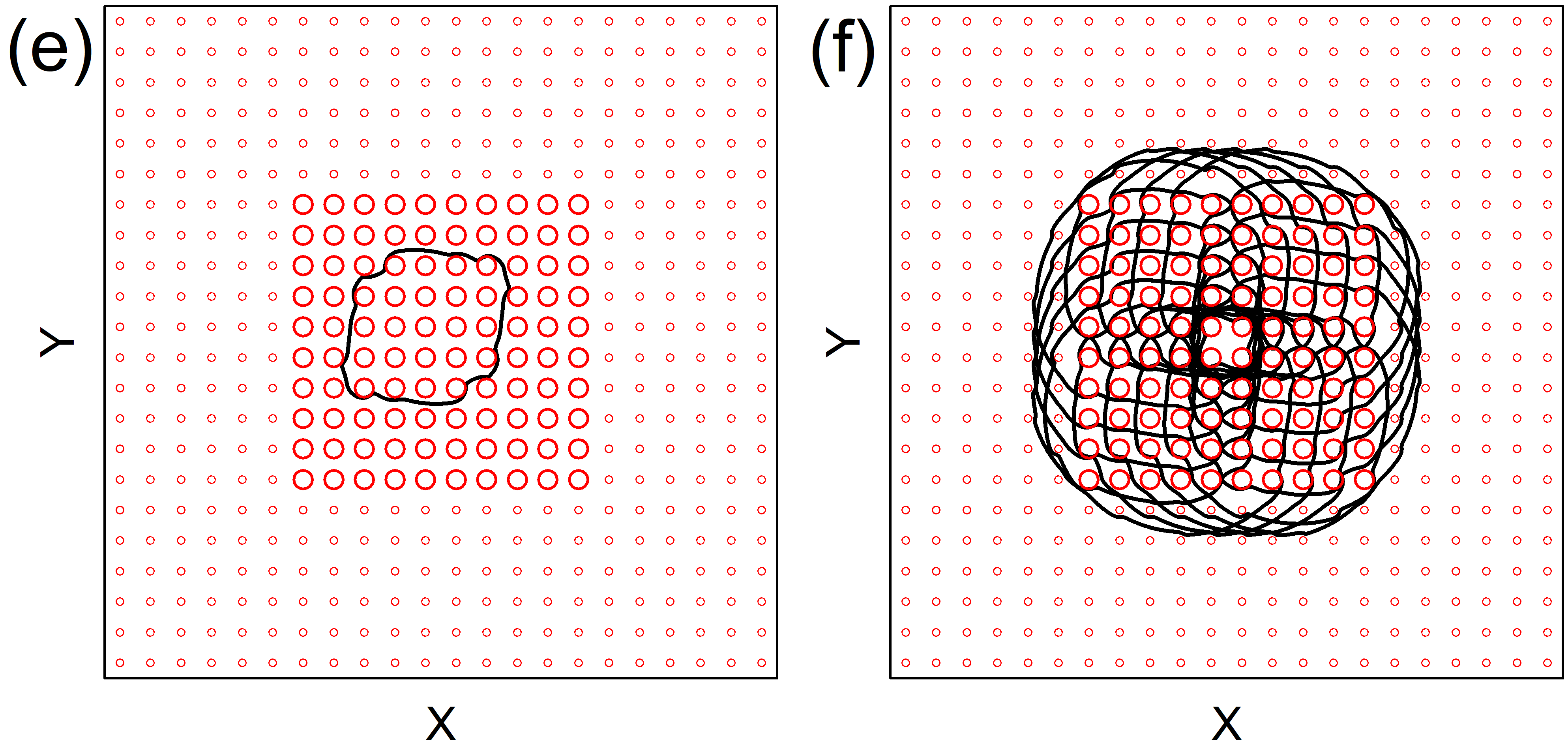}
\end{minipage}
\caption{
Obstacle locations (circles) and skyrmion trajectories (lines)    
for the system in Fig.~\ref{fig:1}
with
$\alpha_{m}/\alpha_{d} = 0.45$ and $\omega = 1\times10^{-5}$. 
(a) A localized orbit at $A = 0.1$ which encircles no obstacles.
(b) Another localized orbit at $A = 0.2$ which encircles a single obstacle.
(c) A localized orbit at $A = 0.4$ in the small obstacle region which
encircles 12 obstacles.
(d) At $A = 0.47$, the skyrmion can translate
along the interface and move around all four corners to form
a large scale clockwise moving orbit. 
(e) At $A = 0.66$, the skyrmion is localized again.
(f) At $A = 0.86$, another translating orbit appears.  
}
\label{fig:6}
\end{figure}

Now that we have established that skyrmions can undergo directed
transport along a straight interface, 
the next question is whether the skyrmion can continue to follow
an interface
that changes direction or has a corner or even multiple corners.
If so,
it would be possible to guide a skyrmion along complex interface
geometries in order to achieve various types of devices. 
To examine this question we consider the geometry illustrated
in Fig.~\ref{fig:1} where the large obstacles are placed in a square
region in the center of the small obstacle lattice.
In Fig.~\ref{fig:6} we show some representative orbits
for the system in Fig.~\ref{fig:1}
with
$\alpha_{m}/\alpha_{d} = 0.45$ and $\omega = 1\times10^{-5}$.
Localized orbits appear in Fig.~\ref{fig:6}(a) at $A=0.1$ and
in Fig.~\ref{fig:6}(b) at $A=0.2$.
A localized orbit forms in the small obstacle region at $A=0.4$
in Fig.~\ref{fig:6}(c).
At $A=0.47$, a sample with a 1D interface has a translating orbit
with motion in the positive $x$ direction. In Fig.~\ref{fig:6}(d),
the skyrmion is able to follow the interface at $A=0.47$ and 
turn at all four corners, forming a large scale
clockwise orbit along which the skyrmion moves
by a distance $2a$ during every drive cycle.
Figure~\ref{fig:6}(e) shows a localized orbit
encircling 23 obstacles at $A = 0.66$.
In Fig.~\ref{fig:6}(f) at $A=0.86$, another large scale
interface-following translating orbit
appears
which is wider than the orbit shown in Fig.~\ref{fig:6}(d).

\begin{figure}
\includegraphics[width=3.5in]{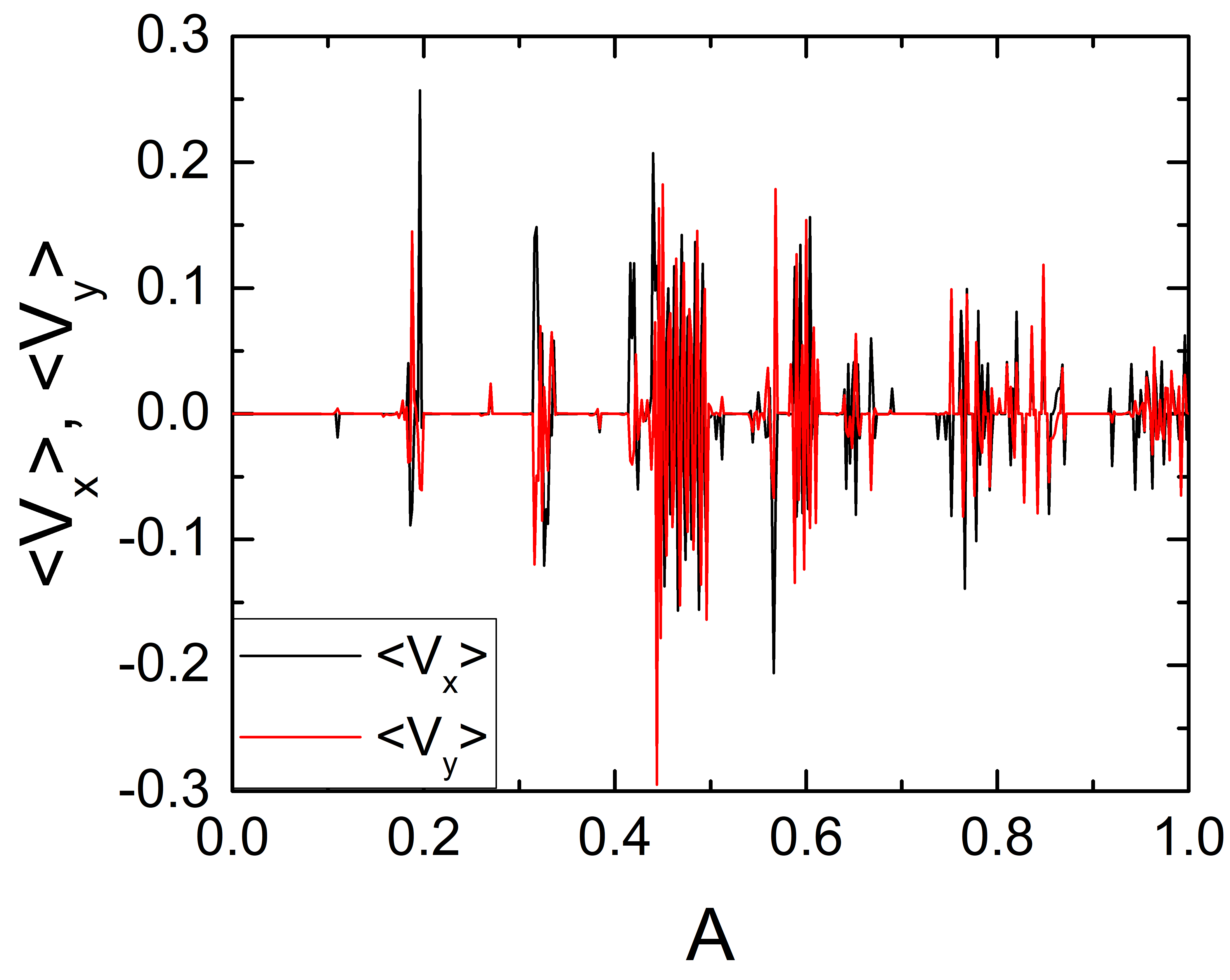}
\caption{ $\langle V_{x}\rangle$ (black)  and $\langle V_{y}\rangle$ (red)
vs $A$ for the system in Fig.~\ref{fig:6}
with
$\alpha_m/\alpha_d=0.45$ and
$\omega=1\times 10^{-5}$.
The regions in which oscillations appear in $\langle V_x\rangle$ and
$\langle V_y\rangle$ correspond to regions in which the skyrmion is
translating around the interface.
  }
\label{fig:7}
\end{figure}

In Fig.~\ref{fig:7} we plot $\langle V_{x}\rangle$ and
$\langle V_{y}\rangle$ for the system in Fig.~\ref{fig:6}.
Unlike the case for motion along a 1D interface,
there are now finite velocity components in 
both directions.
Since
the skyrmion is changing directions as it translates, the velocity 
components do not show the smooth steps found
for the 1D interface motion; however,
the regions where translation
occurs are marked by
large fluctuations or oscillations in $\langle V_x\rangle$
and $\langle V_y\rangle$. 
There are several intervals of $A$
in which 
the skyrmion undergoes transport along the interface.
There are also some intervals in which the skyrmion translates along
straight portions of the interface but cannot turn a corner, so that the
skyrmion motion becomes delocalized once the corner is passed.

\begin{figure}
\includegraphics[width=3.5in]{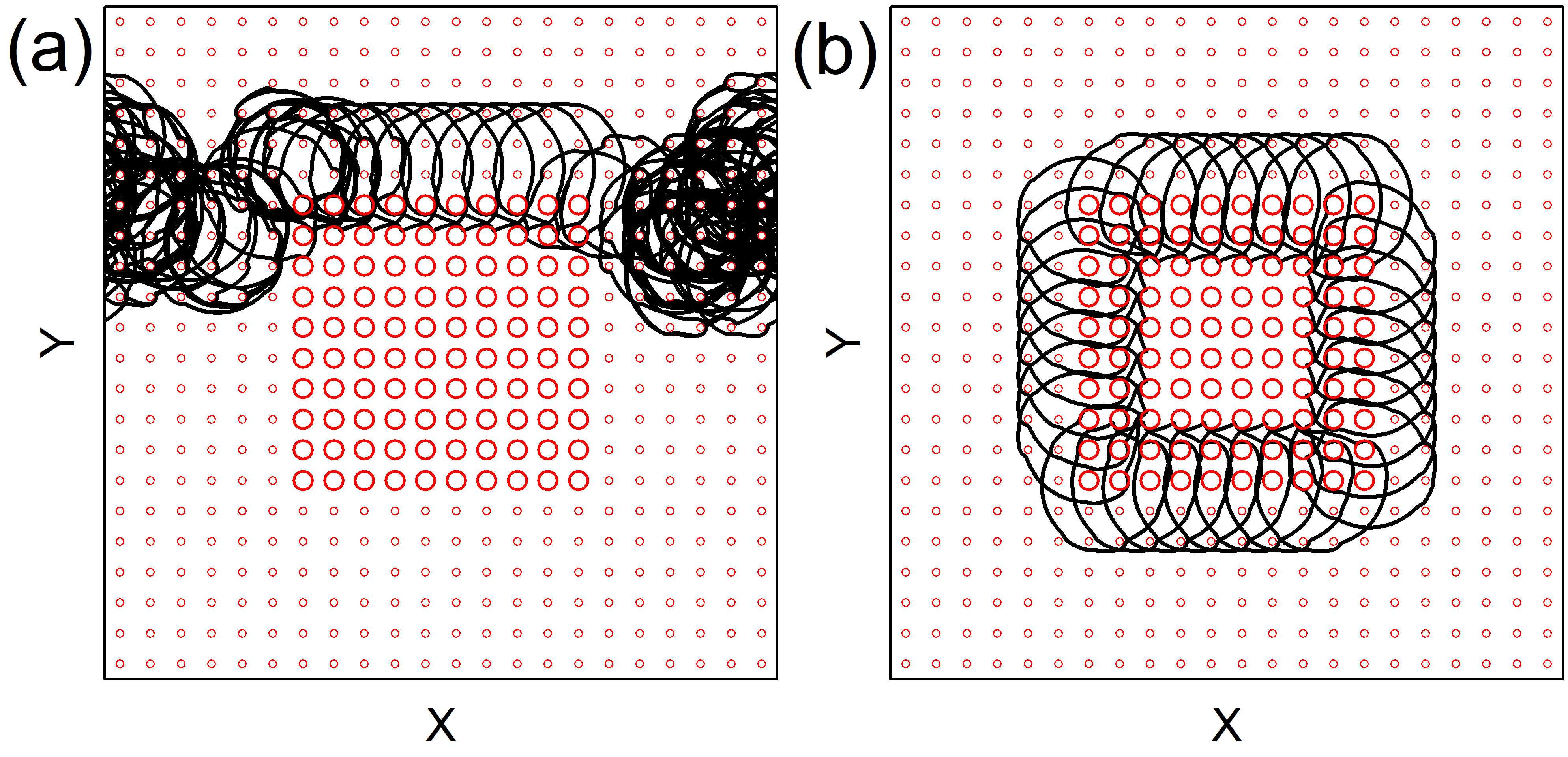}
\caption{Obstacle locations (circles) and skyrmion trajectories (lines)
for a system similar to that in Fig.~\ref{fig:6} with
$\alpha_m/\alpha_d=0.45$
but at $\omega = 2\times 10^{-5}$. (a) At $A = 0.886$, 
the skyrmion can translate linearly along the interface but cannot
turn a corner. (b) At $A=0.895$, the skyrmion can
both translate and turn corners.  
}
\label{fig:9}
\end{figure}

In general, the values of $\alpha_{m}/\alpha_{d}$ and $A$
at which directed motion can occur along a 1D interface are the
same as the values at which
the skyrmion can move along an interface and turn a corner;
however, the interval of $A$ over which the corner-turning behavior
appears is
somewhat
reduced compared to the interval in which 1D translation appears.
We show an example of this in Fig.~\ref{fig:9}(a)
at $A=0.886$ for a system
similar to that in Fig.~\ref{fig:6} but for a higher frequency
of $\omega = 2\times 10^{-5}$.
In this case, the skyrmion motion is chaotic in the region of small
obstacles, but the skyrmion can follow the interface along a straight
line.  It cannot, however, turn the corner, so the motion becomes
delocalized again once the skyrmion reaches the right edge of
the large obstacle region.
Figure~\ref{fig:9}(b) shows the same system at $A = 0.894$ where the skyrmion
can now turn the corners.
As $A$ is increased, we find additional regimes in which the
skyrmion can only move linearly along one side of the interface
before returning to a delocalized or localized state after passing
the corner of the large obstacle region.

\begin{figure}
  \begin{minipage}{3.5in}
\includegraphics[width=3.5in]{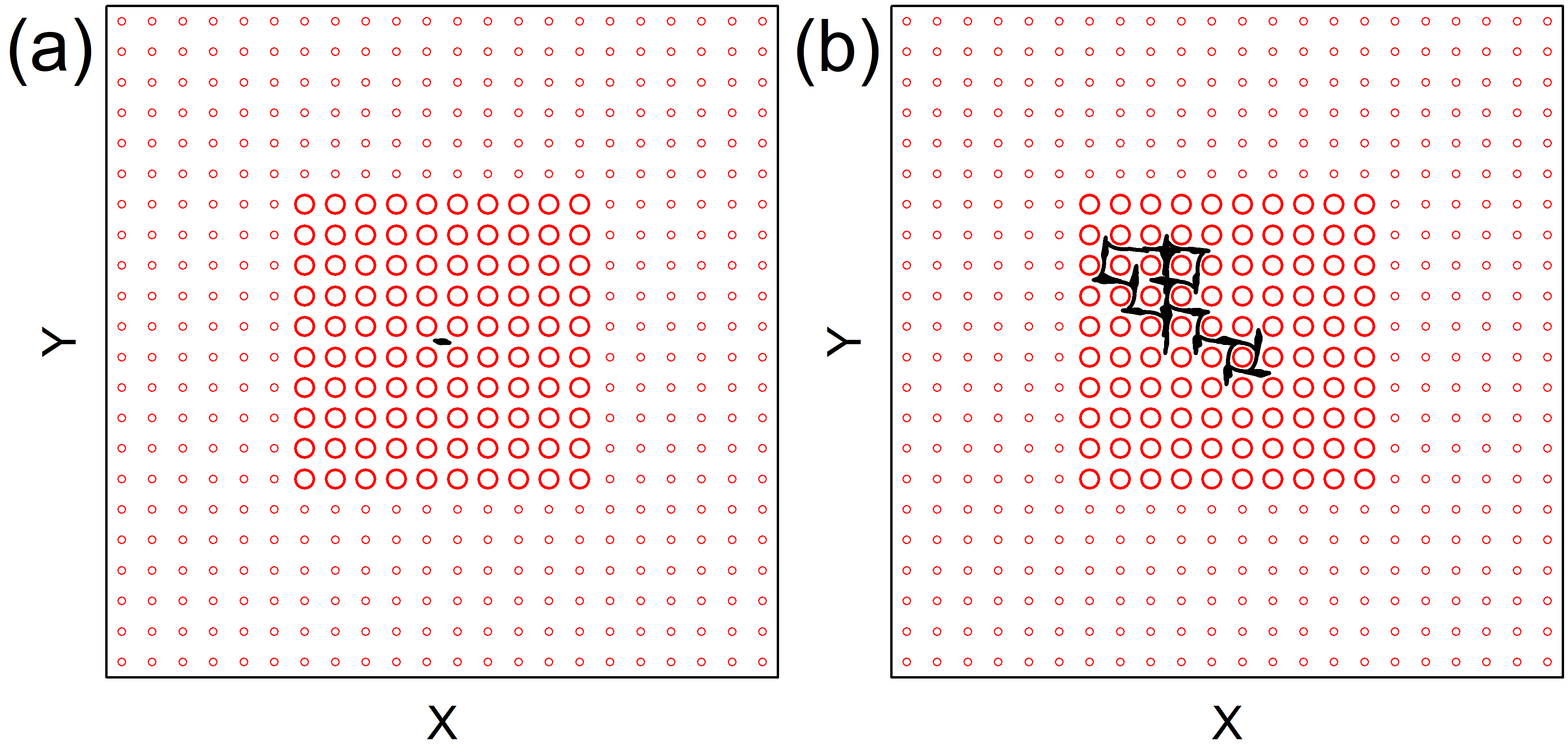}
\includegraphics[width=3.5in]{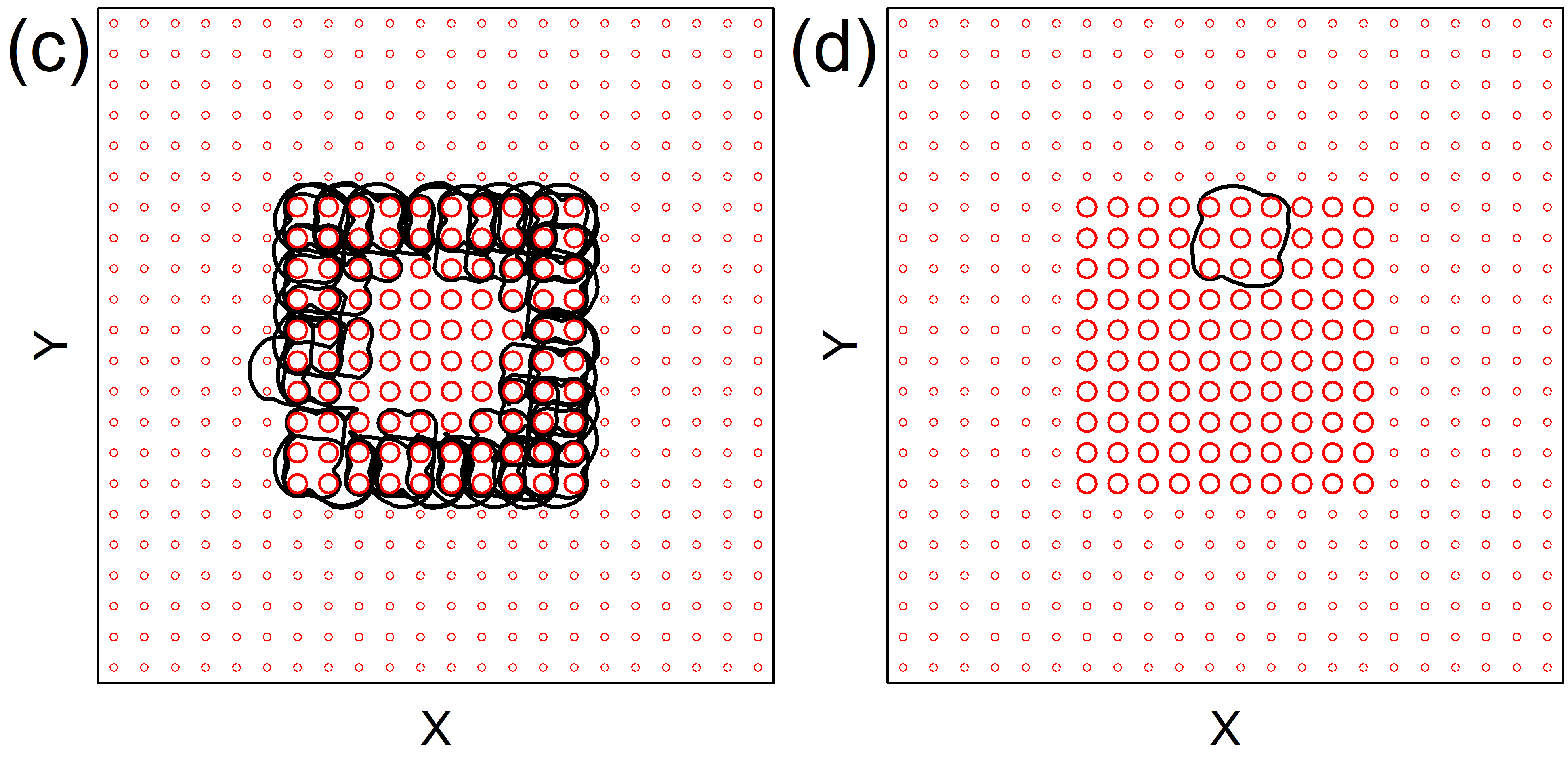}
\includegraphics[width=3.5in]{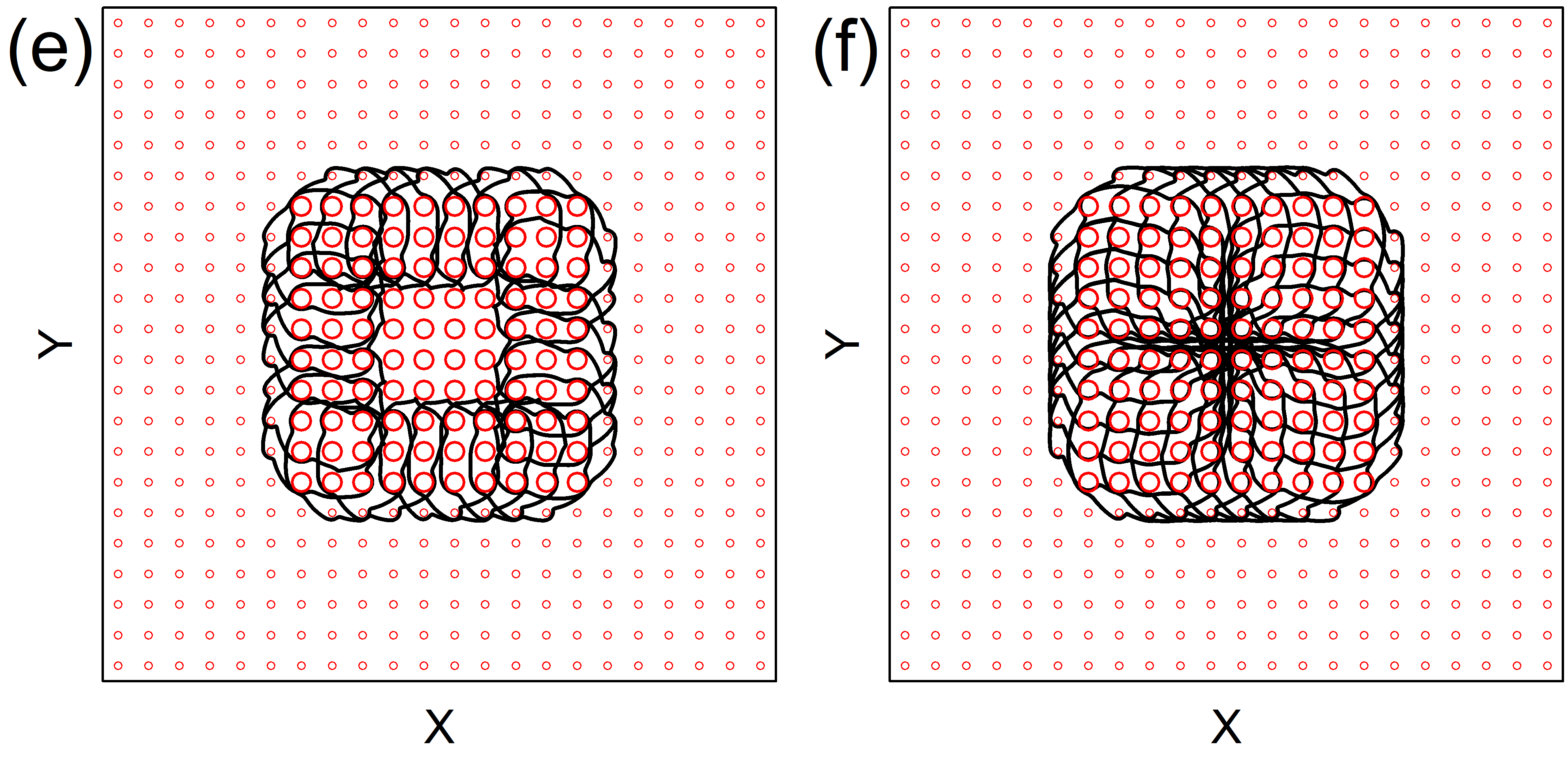}
\includegraphics[width=3.5in]{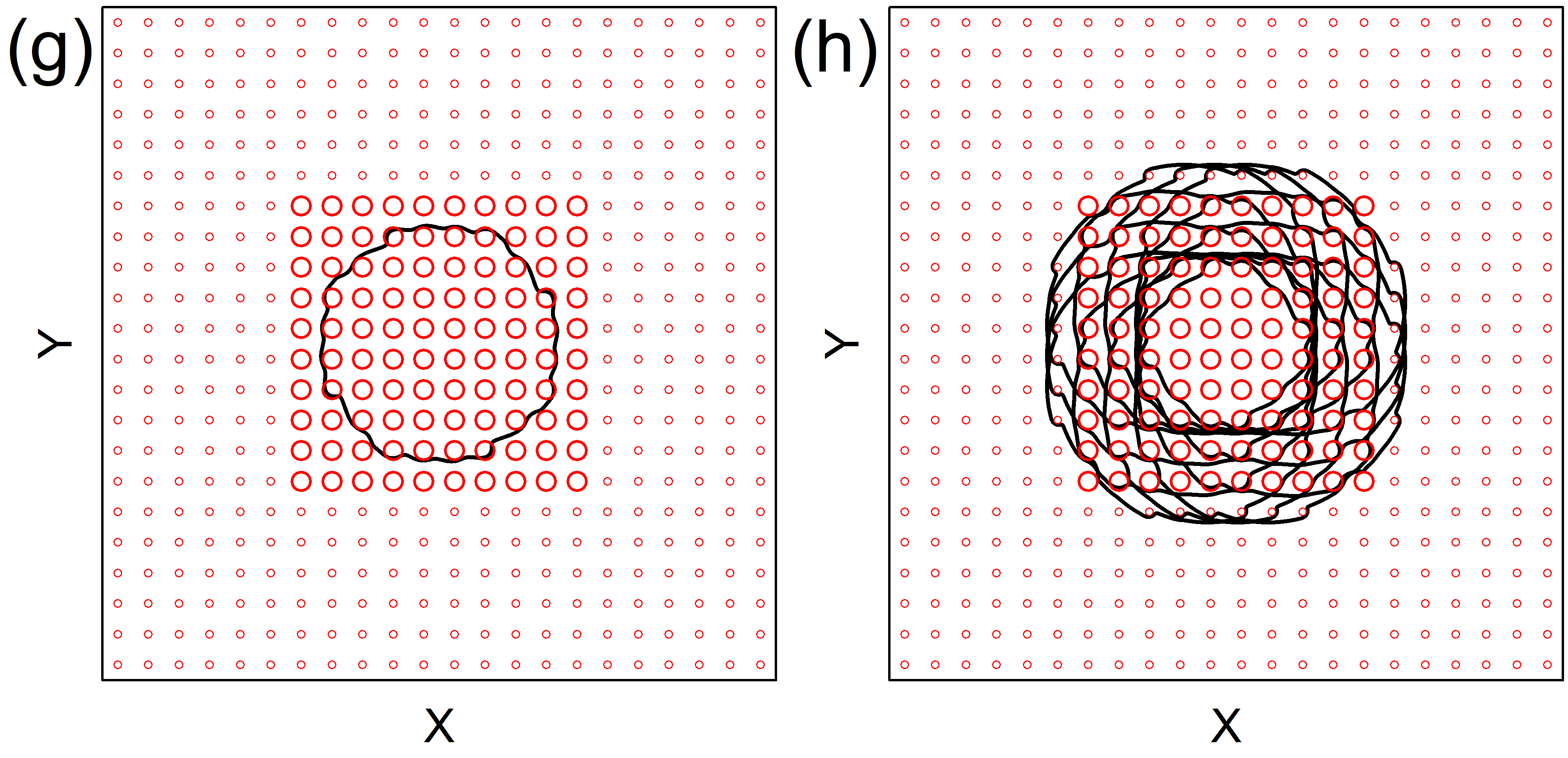}
\end{minipage}
\caption{Obstacle locations (circles) and skyrmion orbits (lines)
for a system with
$\alpha_{m}/\alpha_{d} = 1.732$
and $\omega = 1 \times 10^{-5}$.
(a) Localized motion at $A = 0.06$.
(b) At $A = 0.074$, there is stochastic motion in the bulk but no translation.
(c) At $A = 0.246$ there is interface motion with intermittent chaotic
intervals.
(d) A localized orbit at $A = 0.33$.
(e) A translating orbit at $A = 0.38$.
(f) A translating orbit at $A  = 0.532$.
(g) A localized orbit at $A = 0.74$.
(h) A translating orbit at $A = 0.838$.
}
\label{fig:10}
\end{figure}

For fixed $\omega$, as $\alpha_{m}/\alpha_{d}$ increases
we find a regime in which
the skyrmion can traverse the interface and turn corners
yet has a partially stochastic component to its motion.
In Fig.~\ref{fig:10}(a) we illustrate the localized
orbit that forms for a system with $\alpha_{m}/\alpha_{d} = 1.732$ 
at $A = 0.06$.
In the same system 
at $A = 0.074$, Fig.~\ref{fig:10}(b) shows that the skyrmion undergoes
chaotic motion in the center bulk region but there
is no directed motion.
In Fig.~\ref{fig:10}(c)
at $A = 0.246$,
directed motion occurs along the interface but there are intermittent
windows of chaotic motion,
reducing the efficiency of the transport
to a distance of much less than $a$ per ac drive cycle. 
Figure~\ref{fig:10}(d) shows a localized orbit at $A = 0.33$.
In Fig.~\ref{fig:10}(e), there is a strongly translating orbit
at $A = 0.38$.
Another translating orbit at $A=0.532$ appears
in Fig.~\ref{fig:10}(f).
In Fig.~\ref{fig:10}(g) we illustrate a localized orbit at $A = 0.74$
that encircles 52 obstacles, while Fig.~\ref{fig:10}(h) shows 
another translating orbit at $A = 0.838$. 

\begin{figure}
  \begin{minipage}{3.5in}
\includegraphics[width=3.5in]{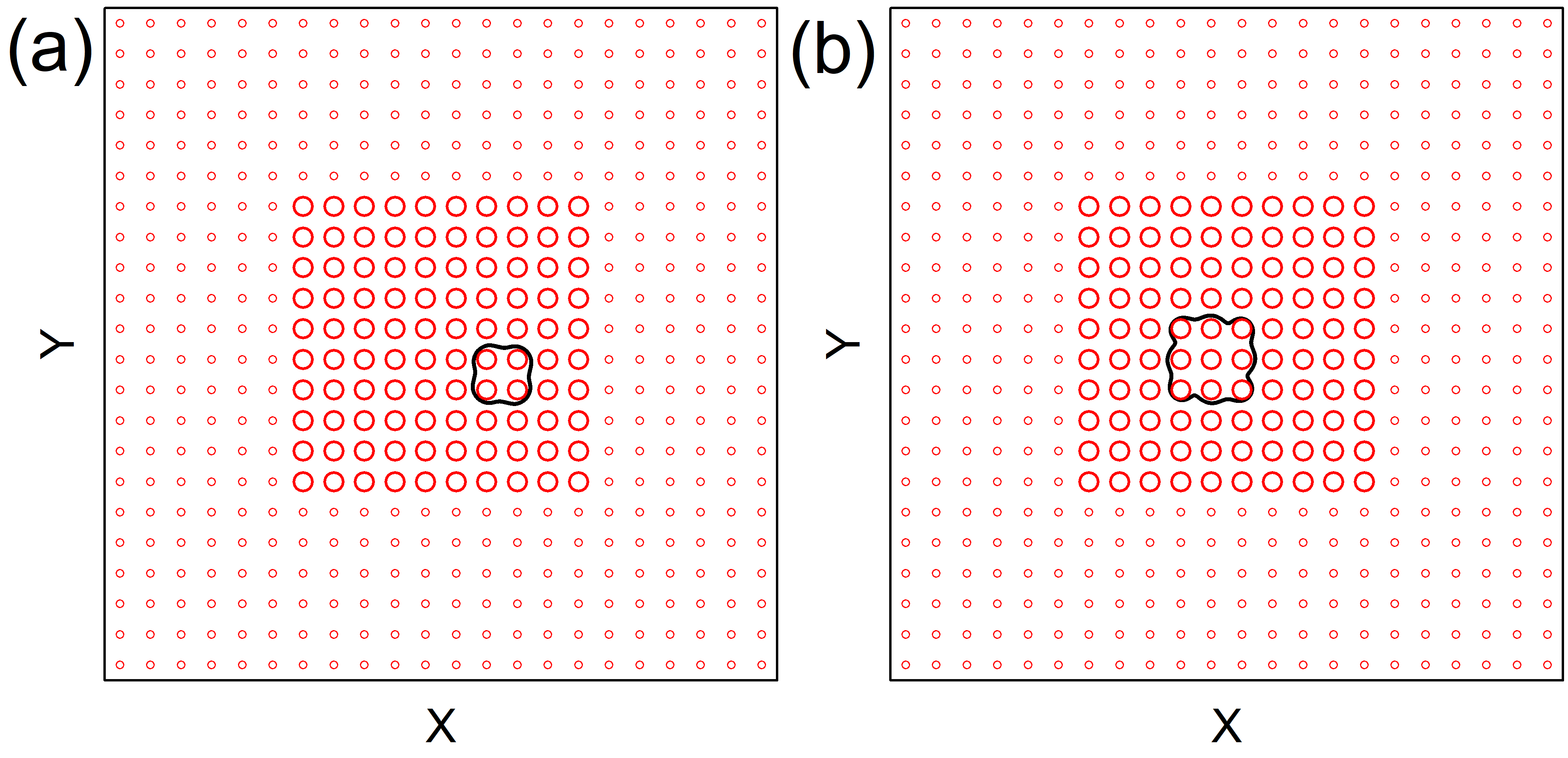}
\includegraphics[width=3.5in]{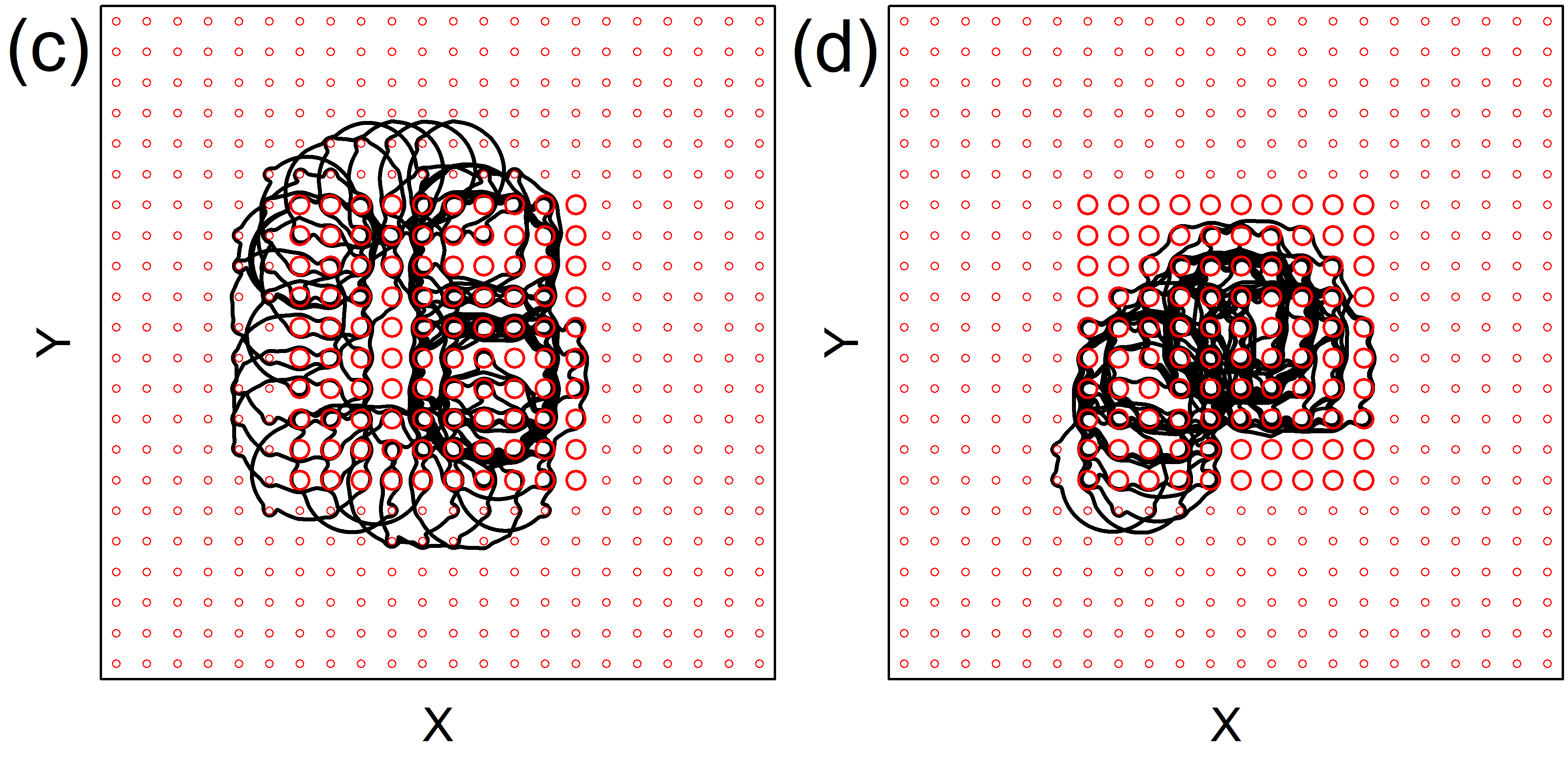}
\end{minipage}
\caption{Obstacle locations (circles) and skyrmion orbits (lines)
for a system with
$\alpha_{m}/\alpha_{d} = 9.962$
and $\omega=1\times 10^{-5}$.
(a) A localized orbit encircling four obstacles at $A = 0.03$.
(b) A localized orbit encircling nine obstacles at $A=0.1$.
(c) At $A = 0.324$, the skyrmion jumps between chaotic motion
in the bulk and translating motion along the interface.
(d) At $A = 0.34$ there is chaotic 
motion in the bulk but no edge transport. 
 }
\label{fig:11}
\end{figure}

\begin{figure}
\includegraphics[width=3.5in]{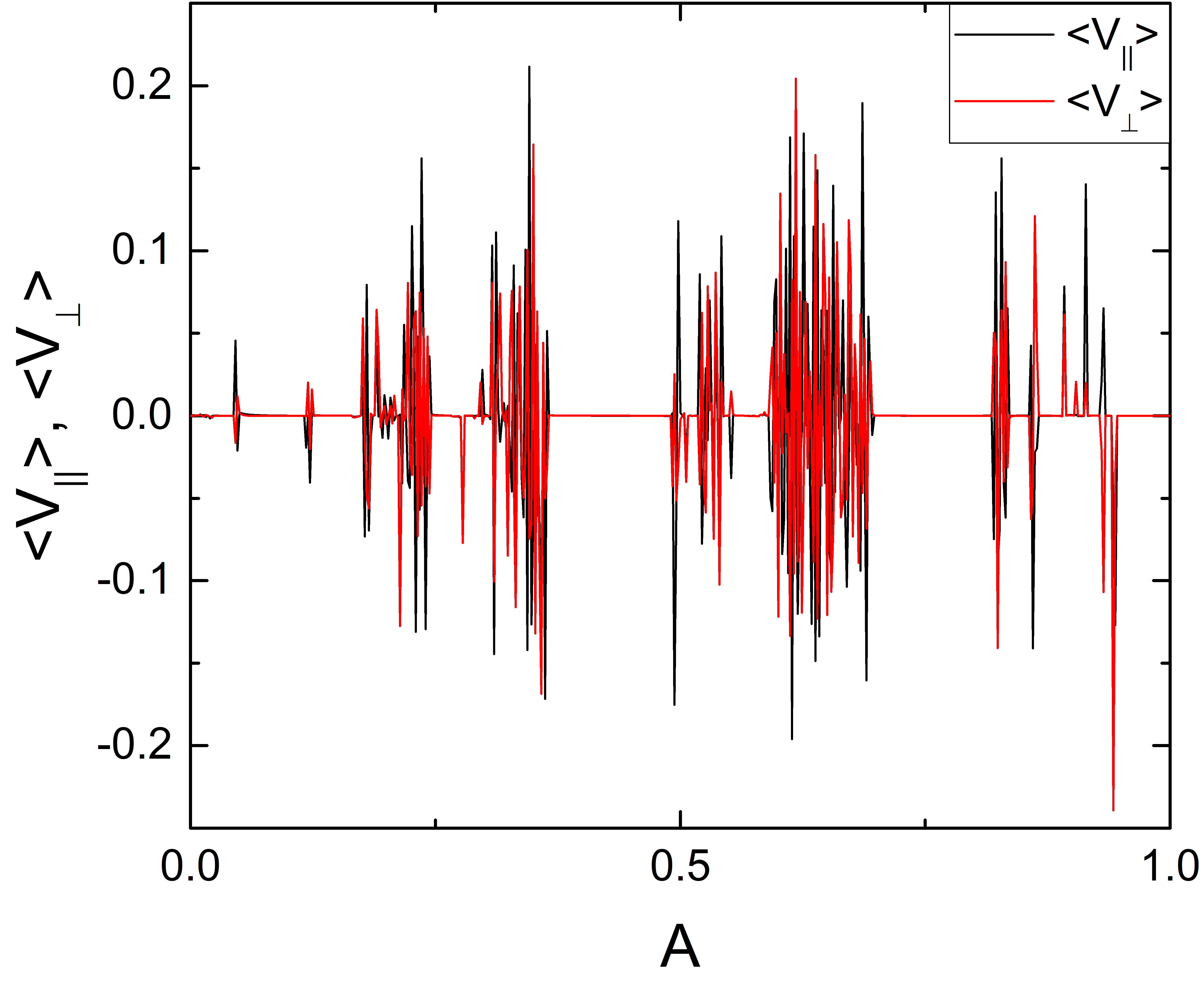}
\caption{ $\langle V_{x}\rangle$ (black) and $\langle V_{y}\rangle$ (red)
vs $A$ for the system in Fig.~\ref{fig:11} with
$\alpha_m/\alpha_d=9.962$ and $\omega=1\times 10^{-5}$. Regions
in which $\langle V_x\rangle$ and $\langle V_y\rangle$ are oscillating
correspond to directed transport around the interface.
 }
\label{fig:12}
\end{figure}

For even higher values of $\alpha_{m}/\alpha_{d}$, chaotic motion
becomes more prevalent.
In Fig.~\ref{fig:11}(a,b) we show the trajectories
for a system similar to that in Fig.~\ref{fig:10} 
but at $\alpha_{m}/\alpha_{d} = 9.962$ 
for $A = 0.03$ and $A=0.1$.
The orbits are localized and the skyrmion
encircles four and nine obstacles, respectively.
For larger $A$, chaotic motion in the bulk accompanies
transport along the edges,
as shown in Fig.~\ref{fig:11}(c) at $A = 0.324$.
Here the 
skyrmion undergoes stochastic motion in the bulk but performs
directed transport along half of the interface
before reentering the bulk region
and returning to a stochastic trajectory.
For long times this pattern repeats itself.
In Fig.~\ref{fig:11}(d) we plot the trajectory at $A = 0.34$ where
the motion is chaotic in the bulk 
but no edge transport occurs along the interface. 
In Fig.~\ref{fig:12} we plot $\langle V_{x}\rangle$ and $\langle V_{y}\rangle$
versus $A$ for the system in Fig.~\ref{fig:11} showing
that there are 
numerous regions where strong oscillations of the velocity
appear,
corresponding to intermittent transport along the interface or ordered
motion in the bulk.
There are several regions of localized motion
in which $\langle V_{x}\rangle = \langle V_y\rangle = 0$.

\section{Discussion}
A recent study
showed directed motion for particles interacting 
with a periodic lattice substrate in what are called
colloidal topological insulators \cite{Loehr18,MassanaCid19}.
Here, a colloid driven in
a circular or closed orbit can exhibit directed
transport when interacting with the interface between two different
types of substrate lattices.
Our results suggest that if a group of skyrmions
is interacting with an interface in a substrate,
various types of edge transport modes could occur
that would create a version of a skyrmion
topological insulator in which
skyrmion motion would not occur in the bulk
but could appear along the edge.
Regarding possible collective effects,
if the number of skyrmions is low we would expect behavior similar
to that described in this work;
however, if 
two or more skyrmions are close together,
they could form a more complex orbit
and move as a group along the edge.
Since the number of degrees of freedom
in such a translating object is higher,
it is possible that a variety of fractional 
motions could occur and that groups of skyrmions could 
form a composite skyrmion state.
We have only considered the case
in which the lattice constant is the same but the size
of the obstacles is different in the two lattices.
It would also be interesting to consider an interface
between two different types of lattices,
such as square and triangular,
or between two lattices with
different lattice constants.
Our particle-based approach
neglects the internal modes of the skyrmions,
and has been shown to produce various types of ratchet effects
for skyrmions interacting with asymmetric substrates
\cite{Reichhardt15aa,Ma17}.
In continuum models
which include skyrmion breathing modes,
new types of ratchet effects can occur even in the absence of
a substrate \cite{Chen19,Chen20}.
This suggests that when internal modes are taken into account,
there could be additional
ways in which the skyrmion could translate along the interface,
and that there could also be additional methods of driving the system such
as with an oscillating magnetic field which could change the
size or shape of the skyrmion as a function of time.   
There is also the question
of thermal effects which could induce phase slips in the directed transport.
Currently the closest geometry
to the system we study
is skyrmions interacting with antidot lattices such as
those recently fabricated in Ref.~\cite{Saha19}.
Using this technique it would be possible to make a sample with a fixed
substrate lattice constant but different antidot sizes in different regions.

\section{Summary}
We have examined skyrmion transport under circular ac driving
along an interface between two 
different obstacle arrays which have the same lattice constant and
different obstacle sizes.
In the absence of an interface,
there is no directed transport.
Inclusion of the interface produces
an additional spatial symmetry breaking
that allows the skyrmion to translate along the interface due to a ratchet
mechanism.
As a result of the periodicity of the lattice,
the skyrmion moves
an integer or rational fractional number of
lattice constants per ac drive cycle.
In addition to linear transport along a 1D interface,
we show that the skyrmion can turn corners
in order to follow the interface and that motion in all four
lattice symmetry directions can be induced with the same ac drive,
suggesting that new types of skyrmion based devices could be created
by having the skyrmion interact with interfaces.
Our results are similar to the motion found for ac driven colloids
in what are called colloidal topological insulators,
where the colloids can undergo transport along 
the interface between two different 
types of substrate lattices.

\subsection*{CRediT authorship contribution statement}
{\bf Nicolas Vizarim}: Methodology, Investigation, Software,
Visualization, Writing - Review \& Editing.
{\bf Charles Reichhardt}: Conceptualization,
Methodology, Writing - Original Draft.
{\bf Pablo Venegas}: Supervision, Funding acquisition.
{\bf Cynthia Reichhardt}: Software, Methodology, Writing - Review \& Editing.

\subsection*{Declaration of Competing Interest}
The authors declare that they have no known competing financial interests
or personal relationships that could have appeared to influence the work
reported in this paper.

\subsection*{Acknowledgement}
This work was supported by the US Department of Energy through
the Los Alamos National Laboratory.  Los Alamos National Laboratory is
operated by Triad National Security, LLC, for the National Nuclear Security
Administration of the U. S. Department of Energy (Contract No.~892333218NCA000001).
N.P.V. acknowledges
funding from
Funda\c{c}\~{a}o de Amparo \`{a} Pesquisa do Estado de S\~{a}o Paulo - FAPESP (Grant 2018/13198-7).

\bibliography{mybib}

\end{document}